\documentclass{article}

\usepackage[english]{babel}

\usepackage[letterpaper,top=2cm,bottom=2cm,left=3cm,right=3cm,marginparwidth=1.75cm]{geometry}

\usepackage{amsmath}
\usepackage{amsfonts} 
\usepackage{graphicx}
\usepackage[colorlinks=true, allcolors=blue]{hyperref}

\title{Lipschitz Extension Initialization for Moving Least Squares
Reconstruction from Sparse Irregular Samples}
\author{Li Chen}

\begin{document}
\maketitle

\begin{abstract}

The idea of using Lipschitz extensions [1,2], or Gradually Varied
Functions (GVFs)[3], for mesh-free scattered data reconstruction was
proposed by the author in 2012 [4]. However, its practical application to
modern mesh-free reconstruction methods has not been fully explored.
Motivated by recent advances in computational tools, including
AI-assisted mathematical programming and software development, we
revisit this idea and investigate the use of a Lipschitz extension as an
initialization step for Moving Least Squares (MLS) reconstruction [5,6]. Our
computational experiments indicate that this initialization
significantly improves the stability and reconstruction accuracy of MLS
under sparse and irregular sampling. This is a preliminary study intended to establish feasibility; a fuller evaluation with additional benchmarks and comparisons is left to future work.
\end{abstract}

\section{ Introduction}

Reconstructing smooth functions from sparse, irregularly sampled data is
a fundamental problem in scientific computing, computer graphics, image
processing, computer-aided geometric design, and data science [7-10]. Widely
used methods such as Moving Least Squares (MLS) produce smooth
approximations when local sampling is sufficiently dense, but their
performance may deteriorate when sample points are sparse or highly
irregular. In this work, we investigate a reconstruction framework that
employs a Lipschitz extension as an initialization step before applying
Moving Least Squares.

The basic idea of using a Lipschitz extension for scattered data
reconstruction is not new. A related approach was proposed by the author
in 2012 (Section 12.2.2 of {[4]}), where Gradually Varied Functions
(GVFs) were suggested as an initial interpolation for mesh-free
scattered data reconstruction. At that time, the computational
investigation was not fully developed. Motivated by recent advances in
computational tools and numerical experimentation, we revisit this idea
and investigate its effectiveness as a preprocessing step for modern
mesh-free reconstruction methods.

Our approach begins by constructing a stable Lipschitz interpolation of
the scattered samples. In the continuous setting, this interpolation is
obtained using the McShane-Whitney extension, a classical Lipschitz
extension theorem. In the discrete setting, the Gradually Varied
Function (GVF) framework may be viewed as a discrete realization of
Lipschitz extensions on graphs and digital domains when the selected
range values satisfy the corresponding Lipschitz condition. Thus, the
McShane-Whitney extension provides a continuous formulation, while GVFs
provide a practical discrete implementation.

The resulting Lipschitz extension serves as a geometric prior that
propagates information into unsampled regions while preserving the
original sample values. Instead of directly applying MLS to sparse
irregular samples, MLS is applied to the initialized field produced by
the Lipschitz extension. This initialization supplies additional
geometric information in regions where local neighborhoods are
insufficiently sampled and therefore improves the stability of the
subsequent local approximation.

Numerical experiments on benchmark scattered data demonstrate that
Lipschitz initialization significantly improves reconstruction quality
under sparse irregular sampling. For a smooth benchmark function sampled
at only one hundred irregular points, the proposed initialization
substantially reduces the reconstruction error compared with direct MLS
while producing a smoother and more stable reconstructed surface. These
experiments suggest that stable Lipschitz extensions provide an
effective initialization for smooth reconstruction from sparse irregular
data and may improve the robustness of existing mesh-free approximation
methods.

The proposed framework does not replace existing reconstruction
algorithms. Instead, it separates the reconstruction process into two
complementary stages: \textbf{(1) stable information propagation through
a Lipschitz extension, and (2) smooth approximation using Moving Least
Squares.} This viewpoint provides a simple, practical, and scalable
framework for scattered data reconstruction and suggests that Lipschitz
extensions may serve as effective geometric priors for a broader class
of mesh-free reconstruction methods.

\section{ Background}

This section briefly reviews the mathematical background of the proposed method, including the McShane-Whitney extension, Gradually Varied Functions (GVFs), and the Moving Least Squares (MLS) approximation. These three components form the basis of the proposed reconstruction framework.

\subsection{ McShane-Whitney Extension}

The Lipschitz extension problem is a classical topic in analysis. 
Let $E\subset \mathbb{R}^n$ be a set of sample points and let

\begin{equation}
f:E\rightarrow\mathbb{R}
\end{equation}

\noindent be a Lipschitz continuous function satisfying

\begin{equation}
|f(x)-f(y)|\le L|x-y|,
\qquad
x,y\in E,
\end{equation}

\noindent where $L$ is a Lipschitz constant.

The McShane-Whitney extension theorem states that $f$ can be extended from $E$ to the entire space ($\mathbb{R}^n$) without increasing the Lipschitz constant. One such extension is the McShane extension.

\begin{equation}
F(x)=
\min_{p\in E}
\{
f(p)+L|x-p|
\},
\end{equation}

\noindent while the Whitney extension is given by

\begin{equation}
F(x)
=
\max_{p\in E}
\{
f(p)-L|x-p|
\}.
\end{equation}

These two formulas provide lower and upper Lipschitz envelopes, respectively. They satisfy

\begin{equation}
F(p)=f(p),
\qquad
p\in E,
\end{equation}

\noindent and preserve the original Lipschitz constant. Because they require only distances to the sample points, they are simple to implement and naturally applicable to scattered data.

In this paper, the McShane extension is used as an initialization step that propagates information from the sampled locations into unsampled regions before local approximation.

\subsection{ Gradually Varied Functions}

Gradually Varied Functions (GVFs) were introduced by Chen as a discrete framework for digital function reconstruction on graphs and discrete domains. Let $D$ denote a graph whose vertices represent discrete sample locations. A function

\begin{equation}
f:D\rightarrow{A_1,A_2,\ldots,A_n}
\end{equation}

\noindent is said to be gradually varied if, whenever two vertices are adjacent, their function values differ by at most one level (meaning that when $x$ and $y$ are adjacent, if  $f(x
)=A_i$ then $f(y)=A_{i-1}, A_i,$ or $A _ {i+1}$).

More generally, when the range values satisfy a Lipschitz condition, GVFs can be viewed as a discrete realization of Lipschitz extensions on graphs. Instead of constructing a continuous function directly, the method propagates values through neighboring vertices while maintaining local consistency.

The existence theorem for gradually varied extensions establishes conditions under which a discrete extension exists. Efficient algorithms have also been developed for constructing such extensions and for refining them iteratively to improve smoothness. The advantage of GVF is that $A_{i}$ can be selected based on actual functions that could be nonlinear to avoid the $L$ value in 
Lipschitz constant is too big. The refinement of $A_{i}$'s (inserting values in between) will provide an exact result to regular 
McShane extension in grid space.

The GVF framework has been successfully applied to digital geometry, image reconstruction, surface fitting, and manifold-valued data. In the present work, it provides the conceptual motivation for using a Lipschitz extension as a preprocessing stage for mesh-free reconstruction.

\subsection{ Moving Least Squares}

Moving Least Squares (MLS) is one of the most widely used mesh-free approximation methods for scattered data reconstruction. Instead of constructing a global interpolating function, MLS computes a local polynomial approximation around each evaluation point.

Given scattered samples

\begin{equation}
{(x_i,f_i)}_{i=1}^{N},
\end{equation}

MLS determines a local polynomial

\begin{equation}
p(x)
\end{equation}

\noindent by minimizing the weighted least-squares functional

\begin{equation}
J(p)
=
\sum_{i=1}^{N}
w(|x-x_i|)
\left(
f_i-p(x_i)
\right)^2,
\end{equation}

\noindent where $w(\cdot)$ is a nonnegative weight function that decreases with distance from the evaluation point. Gaussian kernels and compactly supported weight functions are commonly used in practice.

Because the local approximation changes continuously as the evaluation point moves, MLS generally produces smooth reconstructed surfaces. The method has found numerous applications in computer graphics, mesh-free numerical methods, point-cloud processing, and geometric modeling.

However, the quality of MLS depends strongly on the local distribution of sample points. When the samples are sparse or highly irregular, local neighborhoods may contain insufficient geometric information, leading to unstable approximations and increased reconstruction error.

The objective of this paper is not to modify the MLS algorithm itself. Instead, we investigate whether a stable Lipschitz extension can provide an effective initialization that supplies missing geometric information before MLS performs its local approximation.

\section{Lipschitz Extension Initialization for Moving Least Squares}

The principal contribution of this paper is a two-stage reconstruction framework that combines a Lipschitz extension with the Moving Least Squares (MLS) approximation. Unlike existing MLS variants that modify weight functions, polynomial bases, or neighborhood selection strategies, the proposed method leaves the MLS algorithm unchanged. Instead, it introduces a stable Lipschitz interpolation as a preprocessing step that supplies additional geometric information before local approximation.

\subsection{Motivation}

The quality of an MLS reconstruction depends on the local distribution of sample points. When samples are sufficiently dense and uniformly distributed, neighboring points provide adequate information for local polynomial fitting. However, when the sampling is sparse or highly irregular, large unsampled regions may exist, and the local neighborhood may no longer provide sufficient information for accurate approximation. As a result, the MLS reconstruction may become unstable and exhibit larger reconstruction errors.

The key observation of this work is that these difficulties arise before the MLS approximation itself. Instead of modifying the local fitting procedure, we first construct a globally consistent interpolation that propagates information from the sampled points into the unsampled regions. This interpolated field then serves as a geometric prior for the subsequent MLS reconstruction.

\subsection{Lipschitz Extension Initialization}

Suppose that the scattered sample set is

\begin{equation}
S={(x_i,f_i)}_{i=1}^{N},
\end{equation}

\noindent where

\begin{equation}
x_i\in\mathbb{R}^d.
\end{equation}

Assume that the sampled function satisfies a Lipschitz condition with constant $L$.

For each evaluation point $x$, we first compute a Lipschitz extension using the McShane formula

\begin{equation}
F_M(x)
=
\min_{i}
\{
f_i
+
L|x-x_i|
\}.
\end{equation}

This extension preserves the original sample values while producing a globally defined function satisfying the same Lipschitz bound.

Unlike conventional interpolation methods that rely on triangulations or mesh generation, the McShane extension depends only on distances between points and therefore applies naturally to scattered data in arbitrary dimensions.

The initialized field

\[
F_M(x)
\]

\noindent provides approximate function values throughout the computational domain, including regions where no sample points are available. Consequently, it provides additional geometric information unavailable in direct MLS reconstruction.

\subsection{MLS Reconstruction}

After obtaining the initialized field, the Moving Least Squares approximation is applied in the usual manner.

For an evaluation point $x$, MLS computes a local polynomial.

\[
p(x)
\]

\noindent by minimizing

\begin{equation}
J(p)
=
\sum_{i=1}^{N}
w_i
(
f_i-p(x_i)
)^2,
\end{equation}

\noindent where

\[
w_i=w(|x-x_i|)
\]

is the chosen weight function.

Instead of using only the original scattered samples, the proposed framework augments the reconstruction by incorporating the initialized field generated by the Lipschitz extension. One possible implementation is to minimize:

\begin{equation}
J(p)
=
\sum_i
w_i
(f_i-p(x_i))^2
+
\lambda
\sum_j
v_j
(F_M(y_j)-p(y_j))^2,
\end{equation}

\noindent where
$y_j$ are evaluation points sampled from the initialized field,
$F_M(y_j)$ are the corresponding Lipschitz extension values,
$v_j$ are weighting coefficients, and
$\lambda$ controls the influence of the initialization.

When

\[
\lambda=0,
\]

\noindent the formulation reduces to the classical MLS method.

The additional term introduces prior geometric information into the least-squares fitting while preserving the flexibility of the original MLS formulation.

\subsection{Computational Framework}

The proposed reconstruction procedure consists of four steps.

{\bf Step 1.} Acquire the sparse scattered sample points.

{\bf Step 2.} Construct a Lipschitz extension using the McShane-Whitney extension (or equivalently, a GVF implementation on discrete domains).

{\bf Step 3.} Use the resulting initialized field as additional information for MLS reconstruction.

{\bf Step 4.} Evaluate the reconstructed function over the desired computational domain.

The overall workflow is illustrated conceptually as

\[
\boxed{
\text{Sparse Samples}
\rightarrow
\text{Lipschitz Extension}
\rightarrow
\text{MLS}
\rightarrow
\text{Smooth Reconstruction}
}
\]

This framework separates the reconstruction process into two complementary stages. The Lipschitz extension performs stable information propagation over the entire domain, while MLS performs local smooth approximation. Since the two stages are independent, the initialization strategy can be incorporated into existing MLS implementations with minimal modification.

Furthermore, the same initialization strategy is expected to be applicable to other mesh-free reconstruction techniques, making it a general preprocessing framework rather than a modification of a particular approximation algorithm.

\subsection {Discussion of the method proposed}

We like to revisit the method we discussed in Section 3.3: \\

\noindent \textbf {Definition 3.1 (Lipschitz-Initialized Moving Least Squares, LI-MLS)}

Let

\[
S={(x_i,f_i)}_{i=1}^{N}
\]

be a set of scattered samples, and let

\[
F_M(x)
\]

denote a Lipschitz extension of the sampled data obtained by the McShane-Whitney extension (or equivalently by a Gradually Varied Function implementation on a discrete domain).

A {\textbf Lipschitz-Initialized Moving Least Squares (LI-MLS)} reconstruction is a two-stage reconstruction framework consisting of:

1) {\it Lipschitz Initialization}: Construct a globally defined Lipschitz extension $F_M(x)$ from the scattered samples. The initialized field propagates information into unsampled regions while preserving the original sample values.

2) {\it Moving Least Squares Reconstruction}: Perform the MLS approximation using both the original samples and the initialized field as additional geometric information.

One implementation is obtained by minimizing the weighted least-squares functional,

\begin{equation}
J(p)
=
\sum_i
w_i
\left(
f_i-p(x_i)
\right)^2
+
\lambda
\sum_j
v_j
\left(
F_M(y_j)-p(y_j)
\right)^2,
\end{equation}

\noindent where  $w_i$ and $v_j$ are weight functions;  $y_j$ are points sampled from the initialized field; and $\lambda\ge0$ controls the influence of the Lipschitz initialization.

When

\[
\lambda=0,
\]

LI-MLS reduces to the classical Moving Least Squares method.

The purpose of LI-MLS is not to modify the MLS approximation itself, but to improve its stability under sparse irregular sampling by supplying additional geometric information before local approximation.

\section {Initial Numerical Experiments}

This section evaluates the effectiveness of the proposed Lipschitz-Initialized Moving Least Squares (LI-MLS) framework through numerical experiments. The objective is to investigate whether a Lipschitz extension can provide useful initialization information for MLS reconstruction, particularly when the available samples are sparse and irregularly distributed.

The experiments compare the following reconstruction approaches:

1. {\textbf Direct Moving Least Squares (MLS):}
   The classical MLS reconstruction using only the original scattered sample points.

2. {\textbf McShane-Whitney Lipschitz Extension:}
   The Lipschitz interpolation obtained directly from the sampled data.

3. {\textbf Lipschitz-Initialized Moving Least Squares (LI-MLS):}
   The proposed framework, where the Lipschitz extension is used as an initialization prior before MLS reconstruction.

The reconstruction accuracy is evaluated by comparing the reconstructed function with the original analytical function using the root mean square error (RMSE).

\subsection{Benchmark Function}

To evaluate the reconstruction methods, we first consider a smooth benchmark function defined on a two-dimensional domain.

The test function is selected because it contains both smooth variations and nonlinear behavior, allowing the reconstruction methods to be evaluated under realistic approximation conditions.

The benchmark function is given by

\begin{equation}
f(x,y)=\sin(2\pi x)\cos(2\pi y),
\qquad (x,y)\in[0,1]^2.
\end{equation}

%z = (
%    np.sin(2*np.pi*xy[:,0])
%    *
%    np.cos(2*np.pi*xy[:,1])
%)

The original function is sampled at selected locations, and the reconstructed values are evaluated on a dense grid over the same domain.

Figure 1 shows the original benchmark surface and the corresponding continuous function used for evaluation.

\begin{figure}
\centering
\includegraphics[width=.75\linewidth]{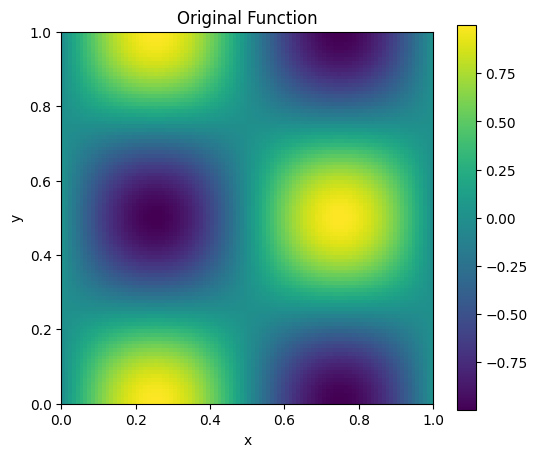}
\caption{\label{fig:Org1} The Original Image.}
\end{figure}

%[Insert Figure 1: Original benchmark function]

The reconstruction error is measured using

\begin{equation}
RMSE
=
\sqrt{
\frac{1}{M}
\sum_{k=1}^{M}
\left(
f(x_k,y_k)-\hat f(x_k,y_k)
\right)^2
},
\end{equation}

where $M$ is the number of evaluation points, $f$ is the original function, and $\hat f$ is the reconstructed function.

\subsection{Sparse Irregular Sampling}

To simulate challenging reconstruction conditions, we randomly select a small number of sample points from the benchmark function while preserving an irregular spatial distribution.

In this experiment, only 100 sample points are retained from the original domain. Compared with dense sampling, this sparse configuration creates large regions with limited local information and provides a challenging test for MLS reconstruction.

Figure 2 illustrates the irregular sample distribution.

\begin{figure}
\centering
\includegraphics[width=.75\linewidth]{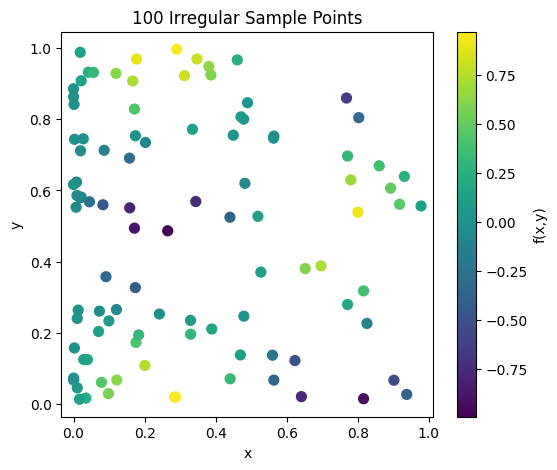}
\caption{\label{fig:C1} The 100 Sample Points.}
\end{figure}

%[Insert Figure 2: Sparse irregular sample points]

The reconstruction process is then performed using:

a) direct MLS from the 100 scattered samples;

\begin{figure}
\centering
\includegraphics[width=.75\linewidth]{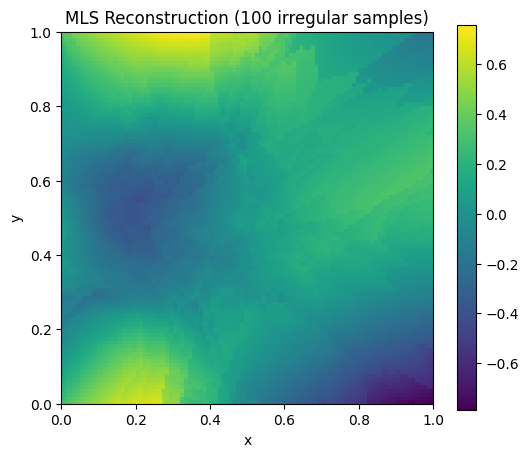}
\caption{\label{fig:C1} The MLS Fitting Result using the 100 Sample Points.  }
\end{figure}

This result is neither a good approximation nor smooth visually. It is caused by the
irregular sample points.

b) McShane-Whitney Lipschitz extension;
 
\begin{figure}
\centering
\includegraphics[width=.75\linewidth]{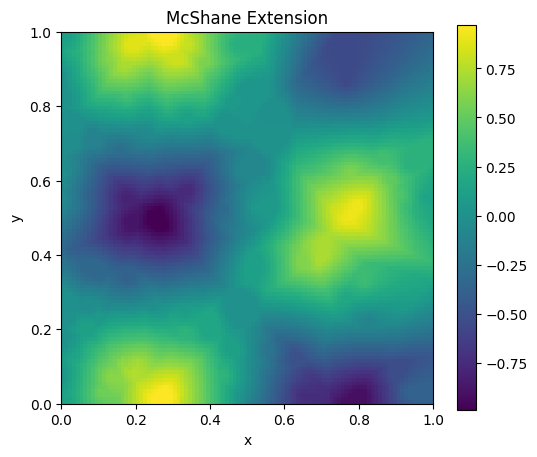}
\caption{\label{fig:C1} McShane Extension using the same 100 samples.  }
\end{figure}

It is a good interpolation compared to Fig 1,  but it is not smooth.

c) LI-MLS using the Lipschitz extension as initialization. (We will talk about not using all interpolated points later.)

\begin{figure}
\centering
\includegraphics[width=.75\linewidth]{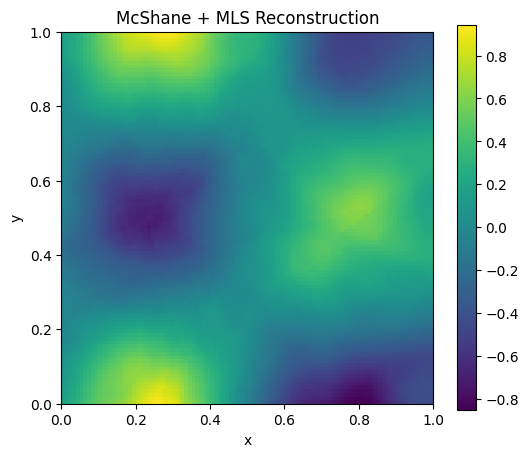}
\caption{\label{fig:C1} Use MLS after McShane Extension.  }
\end{figure}

We can see the results in Fig 5 are smoother than it is in Fig. 4. It is much better
visually compared to Fig. 3. 

The results demonstrate that direct MLS may lose accuracy when the local neighborhoods contain insufficient samples. In contrast, the Lipschitz extension provides a stable interpolation over the entire domain and supplies additional geometric information before MLS fitting.

\subsection{Error Analysis}

%The quantitative reconstruction errors are summarized in Table 1.

%[Insert Table 1: RMSE comparison]

The experiments show that the Lipschitz extension significantly reduces the reconstruction error compared with direct MLS under sparse irregular sampling.

For example, with 100 irregular sample points:

\[
RMSE_{MLS}
=
\text{0.36836517822565207}.
\]

\noindent while

\[
RMSE_{McShane}
=
\text{0.1440843873816341}.
\]

\noindent The LI-MLS reconstruction further improves or maintains the accuracy:

\[
RMSE_{LI-MLS}
=
\text{ 0.20967706631408792}.
\]

Although the McShane extension alone achieves the lowest RMSE, it is only Lipschitz continuous and not smooth — it exhibits kinks/non-differentiable ridges inherent to the min-of-cones construction (visible in Fig. 4). LI-MLS trades a modest increase in RMSE for a smooth, differentiable reconstruction, which is the objective of this study. See Fig. 5.

These results indicate that the Lipschitz initialization provides valuable information in regions where MLS alone lacks sufficient local samples.

The error analysis suggests that the improvement does not come from changing the MLS fitting mechanism itself. Instead, the Lipschitz extension provides a more informative starting field, allowing MLS to perform smoother and more stable local approximation.

\subsection{Selection of Initialization Points}

In the proposed framework, it is not necessary to use every point generated by the Lipschitz extension during the MLS reconstruction stage. Instead, a subset of the interpolated points may be selected according to computational or geometric considerations, such as random sampling, uniform sampling, adaptive sampling, or feature-based sampling. This flexibility can substantially reduce computational cost while preserving the benefits of Lipschitz initialization.

Other methods can also be used for initialization, such as the Voronoi method in [11,12].

\subsection{Discussion}

The numerical experiments demonstrate an important observation:

{\textbf For dense and regular sampling, classical MLS can already provide accurate smooth reconstruction. However, when sampling becomes sparse and irregular, a stable Lipschitz initialization can significantly improve reconstruction quality.}

The role of the two methods is complementary:

a) The Lipschitz extension provides stable interpolation and information propagation.
\newline
b) MLS provides smooth local approximation.

Therefore, the proposed LI-MLS framework separates two different reconstruction tasks:

\[
\text{Information Propagation}
\quad+\quad
\text{Smooth Approximation}
\]

rather than requiring a single method to perform both tasks simultaneously.

The experiments also suggest that Lipschitz extensions may serve as a general geometric prior for other mesh-free reconstruction methods. The current study focuses on MLS as a representative smooth approximation technique; future work will investigate extensions to other reconstruction frameworks and larger-scale scattered datasets.

Overall, these experiments support the hypothesis that a stable Lipschitz extension can improve the robustness of mesh-free reconstruction methods when 
the available samples are sparse and irregular.

\section{Experiments for a Physically Motivated Wave-Field Benchmark}

To further examine the behavior of the proposed Lipschitz initialization
for a different type of smooth field, we consider a physically motivated
wave-field benchmark. Unlike the first benchmark function, which is a
simple analytic test function, the present example contains a localized
oscillatory structure with multiple peaks and valleys. This experiment
provides an additional test of the reconstruction framework under sparse
irregular sampling.

In this example, we use a Gaussian-modulated plane wave of the form

\begin{equation}
u(x,y)
=
A\exp\left[
-\alpha\left((x-x_0)^2+(y-y_0)^2\right)
\right]
\cos(k_xx+k_yy+\phi).
\end{equation}

The Gaussian envelope produces a spatially localized field, while the
cosine term produces multiple oscillatory peaks and valleys. A
time-dependent version can be written as

\begin{equation}
u(x,y,t)
=
A\exp\left[
-\alpha\left((x-x_0)^2+(y-y_0)^2\right)
\right]
\cos(k_xx+k_yy-\omega t+\phi).
\end{equation}

This type of field can be viewed as a localized wave packet or
Gaussian-modulated wave. It provides a physically motivated test for
scattered-data reconstruction because the field contains both localized
structure and oscillatory behavior.

In our numerical experiment, we use

\begin{equation}
u(x,y)
=
\exp\left[
-20\left((x-0.5)^2+(y-0.5)^2\right)
\right]
\cos(12\pi x+8\pi y).
\end{equation}

The same sparse irregular sampling strategy used in the main experiment
is applied to this test function. Since the analytical function is
known, the reconstruction error can again be evaluated directly using
the RMSE.

The following figure shows the original image (Fig. 6.): 

\begin{figure}
\centering
\includegraphics[width=.75\linewidth]{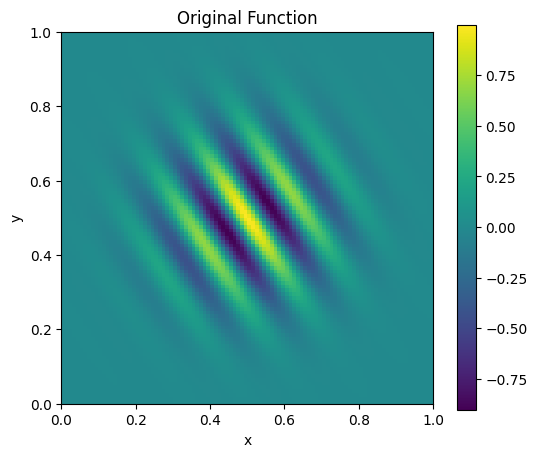}
\caption{\label{fig:C1_smp2} The Original Image.}
\end{figure}

Due to the multiple waves in the center of the picture. One hundred samples used in the 
first experiments are not enough. We now choose 300 samples, as we still want to maintain a lower
number of samples to test our argument in which we said that an initial interpolation would be helpful 
to the final smooth approximation for mesh-free fitting.  Here is Figure (Fig. 7) for the sample points.

\begin{figure}
\centering
\includegraphics[width=.75\linewidth]{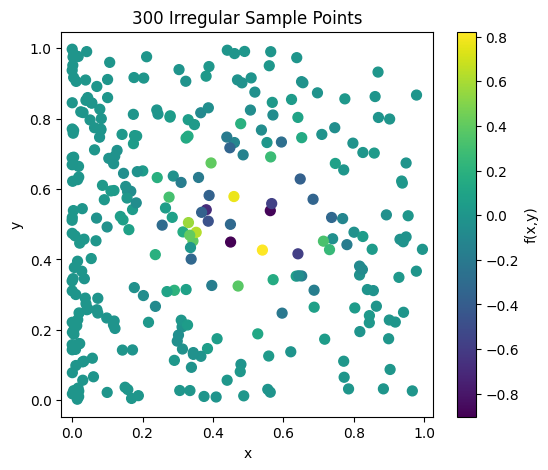}
\caption{\label{fig:C2_smp2} The 300 Sample Points.}
\end{figure}

The following figure (Fig. 8) shows the reconstructed image from 300 sample points using the MLS method: 
\begin{figure}
\centering
\includegraphics[width=.75\linewidth]{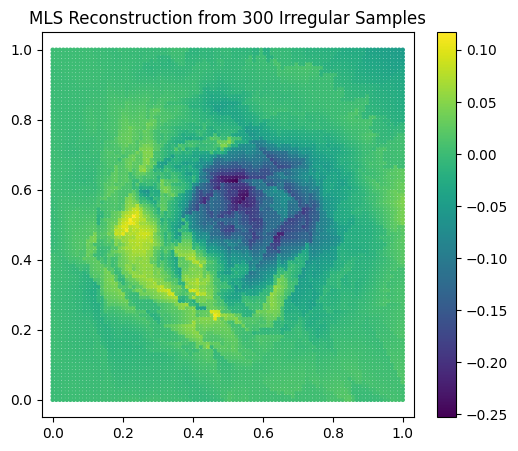}
\caption{\label{fig:C3_smp2} The Reconstructed Image from 300 Sample Points Using MLS.}
\end{figure}

\noindent It is not surprising that the MLS method thinks there are no multiple waves in the center. Rather, 
to do a smooth fitting. There is nothing wrong with interpreting that there is much noise around
 the center of this image.  Just as another example, the reconstructed 
image from McShane Extension using the same 300 samples is shown in the following (Fig. 9.):

\begin{figure}
\centering
\includegraphics[width=.75\linewidth]{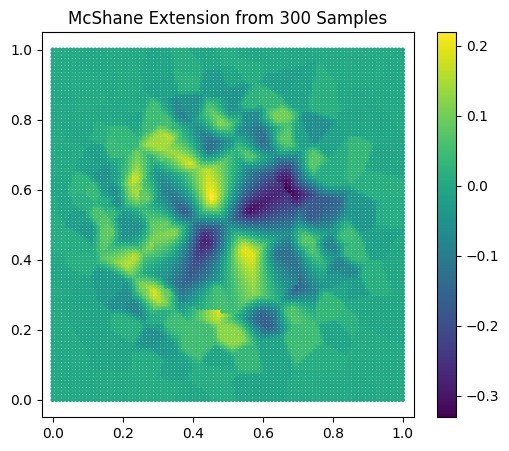}
\caption{\label{fig:C4_smp2} McShane Extension using the same 300 samples.  }
\end{figure}

  This result might indicate that if there is no prior knowledge for a data
set, continuous extension without high capacity of smoothness could be a first choice.
But we do like the following outcome from both methods, see Fig. 10.  

\begin{figure}
\centering
\includegraphics[width=.75\linewidth]{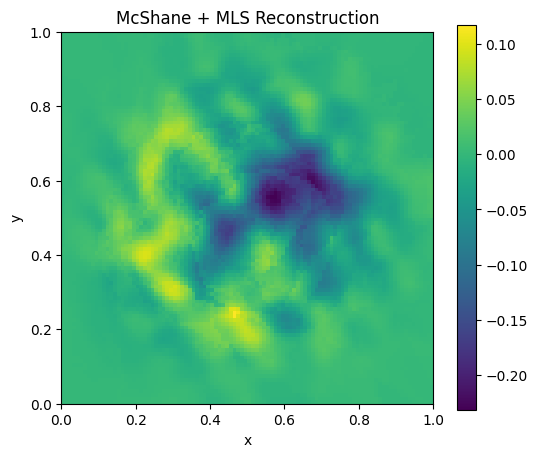}
\caption{\label{fig:C5_smp2} Use MLS after McShane Extension.  }
\end{figure}

The following table shows us terror for different $\lambda$ values. 

\begin{table}[h]
\centering
\begin{tabular}{cc}
\hline
$\lambda$ & RMSE \\
\hline
0.00 & 0.19923283343924836 \\
0.01 & 0.19902457988835945 \\
0.05 & 0.19827358697553626 \\
0.10 & 0.19747906266620220 \\
0.20 & 0.19623435662671615 \\
0.50 & 0.19405946997158777 \\
1.00 & 0.19255818874626200 \\
\hline
\end{tabular}
\caption{RMSE of LI-MLS reconstruction as a function of $\lambda$.}
\label{tab:lambda_rmse}
\end{table}

\section {Conclusions and Future Work}

This paper investigated a new reconstruction framework that uses a Lipschitz extension as an initialization step for Moving Least Squares (MLS) reconstruction from sparse irregular samples. The main motivation is that MLS, although effective for smooth approximation, may suffer from insufficient local information when the sample distribution is sparse or highly nonuniform.

The proposed Lipschitz-Initialized Moving Least Squares (LI-MLS) framework separates the reconstruction process into two complementary stages. First, a stable Lipschitz extension based on the McShane-Whitney extension is constructed from the scattered samples. This stage propagates geometric information into unsampled regions while preserving the original data values. Second, the resulting initialized field is used to support MLS reconstruction, providing additional information for smooth local approximation.

Numerical experiments on a benchmark smooth function demonstrate that the Lipschitz initialization can significantly improve reconstruction quality under sparse irregular sampling. Compared with direct MLS reconstruction, the proposed approach achieves lower reconstruction error and produces a more stable reconstructed surface. These results suggest that Lipschitz extensions can serve as effective geometric priors for mesh-free reconstruction methods.

The contribution of this work is not to replace MLS or introduce a modification of the MLS approximation procedure. Instead, it provides a general preprocessing framework that enhances existing reconstruction methods by improving the information available before approximation. The modular nature of the proposed approach allows it to be integrated with existing mesh-free reconstruction algorithms.

Several directions remain for future investigation. First, larger-scale experiments with different benchmark functions, real-world scattered datasets, and higher-dimensional data will be conducted to further evaluate the scalability and robustness of the method. Second, adaptive strategies for selecting the Lipschitz constant and neighborhood structures will be studied to improve performance for highly nonuniform sampling distributions. Third, the relationship between Lipschitz extensions, Gradually Varied Functions, and higher-order smooth reconstruction frameworks will be further explored.

More broadly, this work suggests that stable Lipschitz extensions provide a useful bridge between interpolation and approximation [13,14]. By first establishing a geometrically consistent field and then applying smooth reconstruction techniques, it may be possible to develop more robust and scalable methods for reconstructing continuous functions from incomplete and irregular data. \\

{\bf Acknowledgments} 

The author gratefully acknowledges the extensive assistance of OpenAI's ChatGPT in mathematical discussions, algorithm prototyping, Python implementation, and manuscript preparation. Anthropic's Claude provided a second review of the manuscript and feedback on the experimental analysis. All mathematical ideas, experimental design, interpretations, and conclusions are the responsibility of the author. The code was placed in Google's Colab.com for experiments.

\end{document}